\documentclass[letterpaper, 10 pt, conference]{ieeeconf}  % Comment this line out
\IEEEoverridecommandlockouts                              % This command is only
\usepackage{graphics} % for pdf, bitmapped graphics files
\usepackage{epsfig} % for postscript graphics files
\usepackage{amsmath} % assumes amsmath package installed
\usepackage{amssymb}  % assumes amsmath package installed

\usepackage{url}
\usepackage[ruled, vlined, linesnumbered]{algorithm2e}
\usepackage{verbatim} 
\usepackage{soul, color}
\usepackage{lmodern}
\usepackage{fancyhdr}
\usepackage[utf8]{inputenc}
\usepackage{fourier} 
\usepackage{array}
\usepackage{makecell}

\SetNlSty{large}{}{:}
\usepackage{graphicx}
\usepackage{amsmath}
\usepackage{array}
\usepackage{booktabs}
\usepackage{tabularx}
\usepackage{adjustbox}
\usepackage{multirow}
\usepackage{hyperref}

\usepackage[style=apa, backend=biber]{biblatex}
\makeatletter

\newcommand{\Rom}[1]{\expandafter\@slowromancap\romannumeral #1@}
\makeatother

\title{\LARGE \bf Vietnamese Legal Information Retrieval in Question-Answering System\\
\large \text{Natural Language Processing - IT4772E} \\
\textbf{Professor: } Le Thanh Huong
}

\author{Nguyen Ba Thiem 20214931% <-this % stops a space 
\\ Doan The Vinh 20210940 \\
Pham Quang Tung 20210919 \\
Tran Van Toan 20214932 \\}

\begin{document}

\maketitle
\thispagestyle{plain}
\pagestyle{plain}

\begin{abstract}

In the modern era of rapidly increasing data volumes, accurately retrieving and recommending relevant documents has become crucial in enhancing the reliability of Question Answering (QA) systems. Recently, Retrieval Augmented Generation (RAG) has gained significant recognition for enhancing the capabilities of large language models (LLMs) by mitigating hallucination issues in QA systems, which is particularly beneficial in the legal domain. Various methods, such as semantic search using dense vector embeddings or a combination of multiple techniques to improve results before feeding them to LLMs, have been proposed. However, these methods often fall short when applied to the Vietnamese language due to several challenges, namely inefficient Vietnamese data processing leading to excessive token length or overly simplistic ensemble techniques that lead to instability and limited improvement. Moreover, a critical issue often overlooked is the ordering of final relevant documents which are used as reference to ensure the accuracy of the answers provided by LLMs. In this report, we introduce our three main modifications taken to address these challenges. First, we explore various practical approaches to data processing to overcome the limitations of the embedding model. Additionally, we enhance Reciprocal Rank Fusion by normalizing order to combine results from keyword and vector searches effectively. We also meticulously re-rank the source pieces of information used by LLMs with Active Retrieval to improve user experience when refining the information generated. In our opinion, this technique can also be considered as a new re-ranking method that might be used in place of the traditional cross encoder. Finally, we integrate these techniques into a comprehensive QA system, significantly improving its performance and reliability.
\\

\end{abstract}


\section{INTRODUCTION}

In today's world of rapidly increasing data volumes, the ability to accurately retrieve and recommend relevant documents has become essential for enhancing the reliability and effectiveness of Question Answering (QA) systems. This is especially true in the legal domain, where precise and contextually appropriate information retrieval is critical. Recent advancements, such as Retrieval Augmented Generation (RAG), have garnered significant attention for their ability to enhance the capabilities of large language models (LLMs) by reducing hallucination issues within QA systems. 

Various methodologies have been proposed to improve document retrieval, including semantic search using dense vector embeddings and the combination of multiple techniques to refine results before they are processed by LLMs. However, these methods often encounter significant challenges when applied to the Vietnamese language. Issues such as inefficient data processing leading to excessive token length in embeddings, extended retrieval times, and overly simplistic result combination techniques leads to instability and limited improvements in retrieval accuracy.

One critical problem that has been largely overlooked is the ordering of the final relevant documents that the LLMs use as references to ensure the accuracy of their answers. Addressing this issue is paramount for improving the overall effectiveness of QA systems in the legal domain.

In this research, we focus on resolving these challenges without delving into the fine-tuning of existing models or the development of new models. Instead, we introduce four key contributions aimed at enhancing Vietnamese legal information retrieval within QA systems. Firstly, we explore practical approaches to data processing to mitigate embedding model inefficiencies. Secondly, we implement data clustering based on similar characteristics to facilitate faster and more accurate retrieval. Thirdly, we propose an enhanced normalization method for Reciprocal Rank Fusion to effectively combine results from both keyword and vector searches. Lastly, we meticulously re-rank the source documents used by LLMs, improving user experience and the precision of the refined information.

These innovative techniques are integrated into a comprehensive QA system, significantly improving its performance and reliability in the Vietnamese legal context. The source code for this project can be found at \href{https://github.com/thiemcun203/Legal-Information-Retrieval.git}{this GitHub repository}.

\section{LITERATURE SURVEY}

In the context of rapidly evolving data volumes, enhancing Question Answering (QA) systems through accurate document retrieval and recommendation has become a focal area of research. The introduction of Retrieval Augmented Generation (RAG) (\cite{Lewis2020RetrievalAugmentedGF}) has been a significant advancement in this domain, particularly in the efforts to mitigate hallucination issues in large language models (LLMs). This section explores related works that have contributed to the development of document retrieval and QA systems, with a brief overview of general advancements. \par

\subsection{Retrieval Augmented Generation (RAG)}
RAG, introduced by \cite{Lewis2020RetrievalAugmentedGF}, combines retrieval mechanisms with generative models to improve the accuracy and reliability of QA systems by reducing hallucinations. This technique leverages external documents to provide more contextually accurate answers, thereby enhancing the performance of LLMs. Subsequent research has built upon this foundation, exploring various retrieval methods and their integration with LLMs to further improve QA system accuracy and reliability. \par

\subsection{Semantic Search and Dense Vector Embeddings}
Semantic search, particularly using dense vector embeddings, has been a prominent method for improving document retrieval. Dense embeddings, represent documents in high-dimensional vector spaces, allowing for more nuanced similarity searches compared to traditional keyword-based methods. The application of BERT \cite{Devlin2019BERTPO} and its derivatives has significantly improved the ability to capture semantic meaning in text, thus enhancing retrieval accuracy.
  
\subsection{Techniques for Improving Retrieval Performance}
Various techniques have been proposed to address the challenges in retrieval performance, notably techniques that utilize multiple retrieval methods (\cite{Fox1993CombinationOM}, \cite{Aslam2001ModelsFM}). One such technique is the Reciprocal Rank Fusion (RRF), introduced by \cite{Cormack2009ReciprocalRF}, which demonstrated the effectiveness of RRF in enhancing retrieval results by combining keyword-based and semantic search methods. However, there still remains room for improvement in how these results are combined and ordered, particularly in the context of QA systems. \par
 
The re-ranking of retrieved documents (\cite{Callan1995SearchingDC}) is another critical area of this project aimed at improving the final output of QA systems. Traditional re-ranking methods, such as those utilizing cross-encoders (\cite{Nogueira2019MultiStageDR}), have shown promise in refining the relevance of documents. These are the essential techniques in information retrieval that have been widely used and proven to improve the relevance and quality of search results.

\section{DATASET CONSTRUCTION}

\subsection{Data Preparation}

At the beginning of the project, our goal was to fine-tune a model for better embedding and re-ranking. However, due to limited training resources and insufficient data, the results were suboptimal. Consequently, we decided to shift to utilizing existing models and optimizing the combination of componenets in our Information Retrieval System. This shift allowed us to focus on preparing data solely for testing purposes. \par

We collected QA pairs from the "Legal Library" website, extracting content into two main parts: user questions and answers provided by lawyers. The questions were meticulously cleaned to retain only the most useful information, removing redundant, duplicate, and greeting words while preserving the main content. Similarly, the answers were carefully cleaned to include only the relevant legal content, ensuring the dataset accurately represents real Vietnamese law. \par

Each dataset was mapped accordingly: the question dataset includes an ID, its content and the relevant document ID, while the law dataset consists of an ID and its content. Each question corresponds to only one relevant document, effectively converting the QA problem into a retrieval problem. \par

% ----- Maybe there can be an image to show an example of the data -----

In the end, we gathered approximately 1,293,347 distinct law articles and 2,081 questions for testing. Below are some statistics of the collected dataset: 

\begin{table}[h!]
\centering
\caption{Distribution of law articles and question Lengths in the Original Legal Document Dataset}
\begin{tabular}{|c|c|c|c|c|}
\hline
\textbf{Length} & \textbf{No.Articles} & \textbf{Percent}  & \textbf{No.Question} & \textbf{Percent}               \\ \hline
$<100$          & 245,001            & 18.95\%   & 1859            & 89.33\%          \\ \hline
101 - 256       & 531,012            & 41.06\%   & 192            & 9.23\%          \\ \hline
257 - 512       & 286,101             & 22.12\%    & 26             & 1.25\%          \\ \hline
513+            & 231,233             & 17.86\%      & 4             & 0.02\%          \\ \hline
\end{tabular}
\end{table}

\subsection{Data Processing}

As shown in the table, many law documents exceed 256 tokens, the input limit for most Vietnamese embedding models, such as the Bi-Encoder from BKAI, which we intend to use. Therefore, we need practical methods to process these data before embedding. We split this intention into two main tasks that can be handled effectively.

Firstly, we split cleaned documents into chunks of fewer than 256 tokens. We prioritize splitting by sections (an article might contain multiple sections) to maintain information units. When section-based splitting isn't feasible, we use an overlap chunking technique, where each partition's beginning tokens are a portion of the previous partition's ending tokens. This method helps retain related information, enhancing the retrieval of relevant pieces. We can also enrich chunks with metadata to add more context. Additionally, summarizing long articles for searching was considered, but it is not suitable for our goal of finding the exact information from documents and might be better for segmentation or clustering problems. For questions longer than 256 tokens, we take the final 256 tokens, as key information usually resides at the end. We also decompose lengthy questions into smaller chunks, search each chunk, and then combine the results for re-ranking.

In the context of our system, to be explained in the next section, a 256-token limit for the Bi-Encoder actually offers an advantage in terms of producing results to the system users. Specifically, the entire chunk can be directly used as a concise reference to back up the answer provided; in contrast, highlighting an entire article, or even just a segment, might still pose too large of text volume for the sake of user experience.

\begin{figure*}[thpb]
      \centering
      \includegraphics[width=0.8\textwidth]{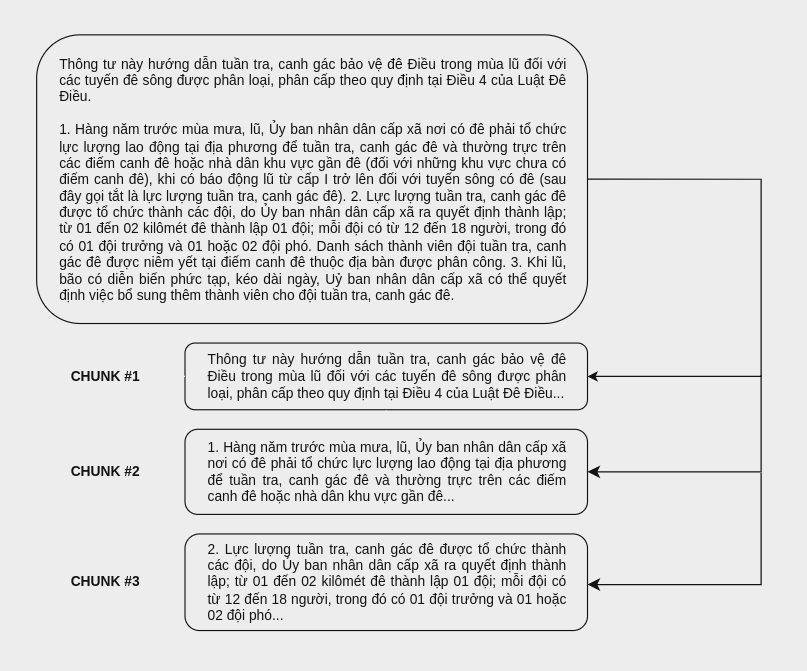}
      \caption{Example of the data preprocessing process, when a large article is split into various overlapping chunks}
      \label{fig:sample_split}
\end{figure*}

Secondly, to enhance embedding vector quality, we segment chunks. Unlike English, Vietnamese is a monosyllabic language, and models perform better when sentences are segmented before tokenization. We chose the RDR segmenter, as the original dataset for training the BKAI model used this segmentation method.

Additionally, for BM25, we remove stopwords to compute accurate scores, avoiding the impact of frequent but less meaningful words. However, for the dense embedding model, we retain words after segmentation before feeding them into the model.

\section{THE SYSTEM ARCHITECTURE}

\begin{figure*}[thpb]
      \centering
      \framebox{\parbox{\textwidth}{\includegraphics[width=\textwidth]{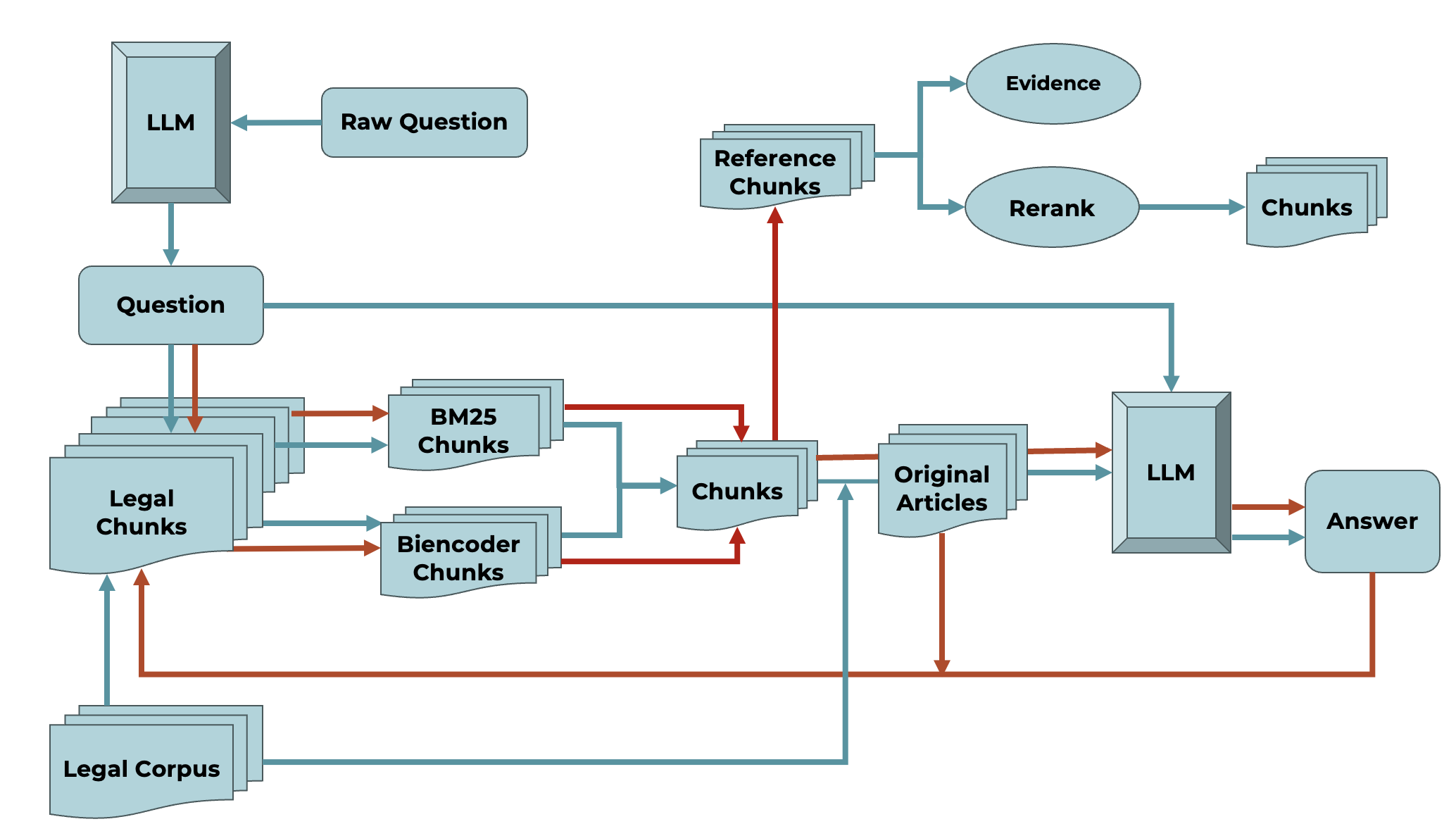}}}
      \caption{ Architecture of our Information Retrieval System as well as Question Answer System}
      \label{fig:4}
\end{figure*}

As an overview, the architecture depicted in Figure 1 comprises a number of main components: a Large Language Model (LLM) in charge of clearly rewriting a user query, a vector database for storing the processed corpus and any other relevant documents, a retrieval mechanism incorporating BM25 search and Dense Vector search utilizing a Bi-Encoder to generate dense embeddings, implemented in an approach that values ranking and re-ranking to return the order of relevancy among chunks of documents, i.e. legal articles. Finally, a component is in charge of generating the final answer and the reference chunks, which we deem to fit as both evidence for the LLM-based answer as well as a re-ranking result which can be utilized by other systems. Details on each component, as well as the flow of our system, will be explained in detail as follows.

\textbf{Vector database for data storage}

Regarding storage needed to save dense embeddings of chunks from legal articles, our project aims to have this data processed and organized in an efficient manner. To achieve this, we utilize the use of vector database which support both BM25 Search and Vector Search for efficient storing and retrieving. Specifically, the vector database support faster searching by using Cosine similarity as a metric and the Hierarchical Navigable Small World algorithm.

\textbf{Rewriting user queries}

\begin{figure*}[thpb]
      \centering
      \includegraphics[width=1\textwidth]{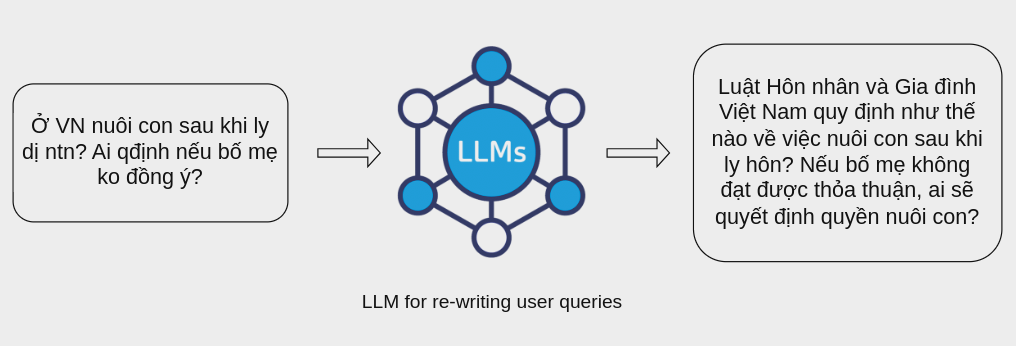}
      \caption{ An example of the Query rewriting module. Notice that the query might contain typos, abbreviations, and vagueness.}
      \label{fig:rewrite_query}
\end{figure*}

Our process begins by optimizing the prompt to instruct the model to generate accurate answers based on relevant retrieved documents and the user's native query. This is often necessary to counteract inadvertent typos and vagueness in the users' questions, which might significantly impact the performance of keyword-based search, and improve clarity. Once the LLM generates a refined version of the query, we use this to search for the actual documents (or article chunks) relevant to the response, essentially verifying the sources used for reasoning. An example can be found in Figure \ref{fig:rewrite_query}. It is to be noted that while the refined query can technically be passed directly to a LLM for an answer or to confirm the existence of a possible answer, general-purpose language models might not be trained on enough niche Vietnamese legal data to fully understand this, which might lead to situations where an answerable question is classified as the opposite. Therefore, we refrain on this practice.

\textbf{BM25 for retrieval}

Regarding keyword search, methods such as TF-IDF and BM25 are commonly used to identify the most relevant documents based on term frequency in both the query and the documents. BM25, represented by the formula below, has a simple yet effective approach that remains widely used today. 

Given a query \(Q\), containing keywords \( q_{1}, \ldots, q_{n} \), the BM25 score of a document \(D\) is:

\[
\text{score}(D, Q) = \sum_{i=1}^{n} \text{IDF}(q_{i}) \cdot \frac{f(q_{i}, D) \cdot (k_{1} + 1)}{f(q_{i}, D) + k_{1} \cdot \left(1 - b + b \cdot \frac{|D|}{\text{avgdl}}\right)}
\]

where \( f(q_{i}, D) \) is the number of times that the keyword \( q_{i} \) occurs in the document \(D\), \( |D| \) is the length of the document \(D\) in words, and \(\text{avgdl}\) is the average document length in the text collection from which documents are drawn. \( k_{1} \) and \( b \) are free parameters. Aftering hyperameter tuning, we choose as \( k_{1} = 1.2 \) and \( b = 0.65 \). \(\text{IDF}(q_{i})\) is the IDF (inverse document frequency) weight of the query term \( q_{i} \). It is usually computed as:

\[
\text{IDF}(q_{i}) = \ln \left( \frac{N - n(q_{i}) + 0.5}{n(q_{i}) + 0.5} + 1 \right)
\]

where \(N\) is the total number of documents in the collection, and \( n(q_{i}) \) is the number of documents containing \( q_{i} \).

However, this method has a significant limitation: it overlooks the semantic similarity between the query and the retrieved documents. In other words, BM25 is only effective when tokens, or words, in the query match exactly against the documents in the corpus, which might not necessarily be the case since there is little control over the way users writes their queries.

\textbf{Vector Search for retrieval}

To address the aforementioned shortcomings of the BM25 search algorithm, dense Vector Search, also known as Semantic Search, has been proposed, particularly to improve search efficiency over methods like cross-encoders. For our implementation, we utilize the best model for Vietnamese embeddings, the Vietnamese Biencoder from BKAI labs, due to the fact that this model boasts an impressive performance when working on legal corpora, such as the Zalo dataset.

\textbf{Combining BM25 Search and Vector Search}

Though in the age of Deep Learning, one might argue that the use of dense embeddings will outperform traditional keyword-based methods; in our project, we would argue that in datasets with lengthy or ambiguous contexts, the encoding model's performance diminishes, and keyword search proves to be a valuable fallback. Subsequently, the scores of the retrieved documents, from both keyword search and semantic search, are normalized to the range [0,1] and combined in a weighted fashion to generate the final results. Therefore, achieving a balance between these methods yields the best efficiency. For balance, we assign a weight of 0.5 to each. An example of the normalizing and combining approach can be seen in Table \ref{tab:original_scores} (original scores) and Table \ref{tab:normalized_scores} (normalized score). Note that the normalized score is calculated as follows.

\begin{equation}
\text{normalized\_score} = \frac{\text{score} - \min(\text{score})}{\max(\text{score}) - \min(\text{score})}
\end{equation}

\begin{table}[h!]
    \caption{Search Results with Original Scores}
    \centering
    \begin{adjustbox}{max width=\columnwidth}
    \begin{tabularx}{\columnwidth}{|l|X|X|X|X|X|}
        \hline
        \textbf{Search Type} & \textbf{(id): score} & \textbf{(id): score} & \textbf{(id): score} & \textbf{(id): score} & \textbf{(id): score} \\
        \hline
        Keyword & (1): 5 & (0): 2.6 & (2): 2.3 & (4): 0.2 & (3): 0.09 \\
        \hline
        Vector  & (2): 0.6 & (4): 0.598 & (0): 0.596 & (1): 0.594 & (3): 0.009 \\
        \hline
    \end{tabularx}
    \end{adjustbox}
    \label{tab:original_scores}
\end{table}

\begin{table}[h!]
    \caption{Search Results with Normalized Scores}
    \centering
    \begin{adjustbox}{max width=\columnwidth}
    \begin{tabularx}{\columnwidth}{|l|X|X|X|X|X|}
        \hline
        \textbf{Search Type} & \textbf{(id): score} & \textbf{(id): score} & \textbf{(id): score} & \textbf{(id): score} & \textbf{(id): score} \\
        \hline
        Keyword & (1): 1.0 & (0): 0.511 & (2): 0.450 & (4): 0.022 & (3): 0.0 \\
        \hline
        Vector  & (2): 1.0 & (4): 0.996 & (0): 0.993 & (1): 0.986 & (3): 0.0 \\
        \hline
    \end{tabularx}
    \end{adjustbox}
    \label{tab:normalized_scores}
\end{table}

Up until this point, the system is working on chunks rather than full articles. After the combined scores have been obtained, we track back to the original articles containing these chunks to be passed to a Large Language Model to extract answers in a subsequent step. This decision is made so as to minimize the loss of context from only using the chunks. Note that the ordering of articles here is also based on score-based ordering of the chunks. For example, if results after combining BM25 and Vector Search, in descending order, are:

\[
\text{chunk\_A} \rightarrow \text{chunk\_B} \rightarrow \text{chunk\_C} \rightarrow \text{chunk\_D}
\]

where chunk\_A and chunk\_C come from \textit{article\_1}, chunk\_B from \textit{article\_3}, and chunk\_D from \textit{article\_2}, the resulting article ordering will be:

\[
\text{article\_1} \rightarrow \text{article\_3} \rightarrow \text{article\_2}
\]

\textbf{LLM-powered answer generation and Active Retrieval}

At this step, the aforementioned retrieved documents are passed to a LLM to leverage its power in understanding the context and extracting the correct spans of text containing the information relevant to the user query and generating an answer. Here, we also specify that the LLM should be willing to acknowledge its inability to answer a question based on proprietary documents. In addition, since the input token length of LLMs are also limited, one prompt to the language model might only be able to handle 2 to 5 legal articles. In the case the number of retrieved documents from the previous step exceeds this capacity and the LLM is consistently not able to answer the query, the same iteration will be applied on the next subset of articles. Once all of the documents have been exhausted and the LLM is still adamant that there exists no information to answer, a process called Active Retrieval takes effect.

In Active Retrieval, the system will explore even the lower-rank docs that are scored using the previous combination of BM25 and Vector Search. The chunks will be converted back into original documents, which are then passed to the LLM to generate answers, in a similar fashion to above. In this way, the system can attempt to its fullest extent to find an answer; only upon reaching a limit to the number of chunks to consider will this practice be terminated - in that case, a definite "No answer found" will be returned.

The optimized prompt used in this system is as follows: \par

\textit{You are an expert lawyer in Vietnam, tasked with answering frequently asked questions (FAQs) from customers about Vietnamese law based on the given information.
Please use, gather, and deduce based on the knowledge in the following information to answer the user's question.
Please respond accurately, concisely, and to the point, without being too verbose. \\
Relevant legal information: [Retrieved Legal Articles] \\ 
User's question: [Question]\\
}

\textbf{Re-ranking process}

Assuming that the previous component of the system runs smoothly, the re-ranking process takes place as we use the LLM-powered answer to re-rank the chunks that have been selected by the BM25-Vector Search combination, reverted back to original documents, and influenced the answer. Specifically, these chunks are re-ranked based on a combination of BM25 and Vector Search again, but by comparing them to the LLM-generated answer. Similar to the explanation above, this combination returns a score that can be used to re-rank the chunks. Afterwards, the chunks can be used as direct references that support the answer to the user query, or they might be used as an intermediate result for yet another re-ranking attempt. While our system does not implement a third re-ranking, the format of the output facilitates this idea - in this case, we recommend that all chunks from the top specified k documents, not just chunks from articles that produce answers, are re-ranked based on the LLM answer.

In the case where the LLM-powered answer is "No answer found", the original corpus, splited into chunks, has already had a negative sentence readily embedded; as a result, when all chunks are re-ranked based on the answer and the combination of BM25 and Vector Search, this specific embedding will most likely be guaranteed to be at the top, signalling the system not to offer any reference when attempts to answer fail.

Overall, the goal behind this loop-like system is to add complexity to reordering of chunks and have them be more accurately relevant, by directly leveraging the output answer. In serving as a re-ranking approach, this can be an alternative to the traditional cross-encoder, which tends to be slower and biased, especially with the increased volume and diversity of data today. In Figure \ref{fig:4}, this process is shown using red arrows.

\textbf{Large Language Model used}
For components requiring the extensive abilities of LLMs, we utilize the latest open-sourced LLaMA 3 70B model released by Meta AI, which demonstrates impressive performance in reasoning and answering questions, comparable to OpenAI's GPT-4.

\section{EXPERIMENTS}

To accurately analyze the performance of our system, we will compare the results obtained with each component that makes up our extensive system, under various settings and data inputs. To prevent overfitting, in addition to the Zalo Legal Dataset, we use our own independently crawled dataset, strictly adhering to the aforementioned standards of data cleaning and preprocessing. We then perform tests on the original dataset as well as on datasets that have been processed through chunking and enriched with metadata.

The components we evaluate include:

1. Only using the BM25 algorithm for ranking relevancy.

2. Only using cosine similarity between the Bi-Encoder's embeddings of a query and document chunks.

3. A hybrid search scheme combining the results of both BM25 and Bi-Encoder.

4. A hybrid architecture incorporating the aforementioned LLM-answer-based re-ranker.

The results of these experiments are presented in Table IV, offering a comprehensive comparison of the different models' performance across the various data settings.

\begin{table*}[ht]
\centering
\begin{tabular}{|l|l|c|c|c|c|}
\hline
\textbf{Dataset} & \textbf{Metrics} & \textbf{BM25} & \textbf{Biencoder} & \textbf{Hybrid} & \textbf{LLM Reranker} \\ \hline
\multirow{3}{*}{Original Data} & Recall@1 & 0.5 & 0.6 & 0.78 & 0.8 \\ \cline{2-6} 
                               & Recall@10 & 0.7 & 0.8 & 0.85 & 0.87 \\ \cline{2-6} 
                               & Recall@100 & 0.8 & 0.88 & 0.9 & 0.9 \\ \hline
\multirow{3}{*}{Processed Data} & Recall@1 & 0.65 & 0.74 & 0.78 & 0.86 \\ \cline{2-6} 
                               & Recall@10 & 0.73 & 0.93 & 0.95 & 0.98 \\ \cline{2-6} 
                               & Recall@100 & 0.89 & 0.97 & 0.98 & 0.98 \\ \hline
\end{tabular}
\caption{Comparison of Model Performance on Different Datasets}
\label{table:comparison}
\end{table*}

\begin{table*}[ht]
\centering
\begin{tabular}{|l|c|c|c|c|}
\hline
Metric & BM25 only & Bi-Encoder only & Hybrid Search & Hybrid with Active Retrieval \\ \hline
Recall@1 & 0.65 & 0.74 & 0.78 & 0.86 \\ \hline
MAP@1 & 0.65 & 0.74 & 0.78 & 0.86 \\ \hline
Recall@10 & 0.73 & 0.93 & 0.95 & 0.98 \\ \hline
MAP@10 & 0.72 & 0.92 & 0.93 & 0.98 \\ \hline
\end{tabular}
\caption{Comparison of Model Performance on Different Datasets - on Recall \& MAP@K}
\label{tab:comparison_2}
\end{table*}

To analyze the performance, we compare the different components in different settings of metrics and test data. Specifically, we test by using both the original and unprocessed dataset, along with the data that has been enhanced with overlapping chunking and added metadata.

From Table IV, it can be observed that data that has undergone appropriate transformations (removing stop words, overlapped chunking to increase the number of chunks and widen context, adding metadata such as article name, etc.) produce significantly better metrics. This reinforces the importance of careful data preparation for NLP tasks in general, and for such a niche domain as Vietnamese law in specific.

In addition, we can see that there is a sharp increase in performance, proportional to the rise in the complexity of the included components. Specifically, the BM25 algorithm depends heavily on the similarity in words between the query and the corpus, and because Vietnamese is a highly diverse, where the same idea can be expressed in numerous manners, this approach might not be able to fully translate the essential semantic meaning of the question into an answer.

By contrast, transformer-based models have recently emerged as a powerful approach to fully capturing the context and meaning of natural language, significantly traditional methods, even in the task of information retrieval where BM25 is a long-standing state-of-the-art solution. This is also reflected in Table IV - the recall for all k's (k = 1, 10, 100) rise compared to BM25. Another interesting insight, albeit simple, is the fact that recall always increase as k increases. From a user experience perspective, returning too many results defeats the purpose of question-answering, yet returning too few places the system at risks of inaccuracy and hallucination.

A surprising empirical conclusion lies in the fact that a composition of keyword-based search (BM25) and semantic search (via embeddings), which is used in our system, comprises an ensemble approach that actually outperforms the individual components. Specifically, this approach works by allowing the Bi-Encoder to re-rank the results from BM25, which offers both a boost in computation efficiency, as well as leveraging the strengths of both methods.

Finally, the entire system, namely the LLM Re-ranker we introduce in this report, boasts the best performance, gaining up to 0.98 recall on the processed dataset. Once again, this confirms the potential of incorporating LLM in a Retrieval-Augmented Generation system.

\section{CONCLUSIONS}
This report proposes a set of modifications to address limitations of the Retrieval-Augmented Generation (RAG) system for Vietnamese Question Answering (QA). By tackling challenges like inefficient data processing and document ordering, these modifications aim to improve the overall performance and reliability of the system. The proposed techniques, including Vietnamese-specific data processing, enhanced fusion methods, and active information re-ranking, have the potential to significantly benefit Vietnamese QA systems, particularly by mitigating hallucination issues in large language models.  The active retrieval method, specifically,  might even offer a new approach to re-ranking documents, potentially replacing traditional cross-encoders. Integrating these advancements into a comprehensive QA system holds promise for a more robust and reliable Vietnamese language information retrieval tool.

\addtolength{\textheight}{-12cm}   % This command serves to balance the column lengths

%%%%%%%%%%%%%%%%%%%%%%%%%%%%%%%%%%%%%%%%%%%%%%%%%%%%%%%%%%%%%%%%%%%%%%%%%%%%%%%%

%%%%%%%%%%%%%%%%%%%%%%%%%%%%%%%%%%%%%%%%%%%%%%%%%%%%%%%%%%%%%%%%%%%%%%%%%%%%%%%%

%%%%%%%%%%%%%%%%%%%%%%%%%%%%%%%%%%%%%%%%%%%%%%%%%%%%%%%%%%%%%%%%%%%%%%%%%%%%%%%%
\section*{ACKNOWLEDGMENT}

We would like to express our deepest gratitude towards Prof. Le Thanh Huong for her valuable comments, guidance, and suggestions to improve the quality of the work throughout the project.

%%%%%%%%%%%%%%%%%%%%%%%%%%%%%%%%%%%%%%%%%%%%%%%%%%%%%%%%%%%%%%%%%%%%%%%%%%%%%%%%
\printbibliography

\end{document}